# Balanced activation in a simple embodied neural simulation

Authors: Peter J. Hellyer [1,2,3], Claudia Clopath [1], Angie A. Kehagia[2], Federico E. Turkheimer [2], Robert Leech[3]

1. Department of Bioengineering, Imperial College London, Exhibition Road, London, SW7 8TZ, UK
2. Centre for Neuroimaging Sciences, Institute of Psychiatry, Psychology and Neuroscience, King's College London. De Crespigny Park, London, SE5 8AF, UK
3. Computational, Cognitive and Clinical Neuroimaging Laboratory (C[3]NL), Imperial College London, Hammersmith Hospital, Du Cane Road, London, W12 0NN, UK

Corresponding author: Robert Leech (r.leech@imperial.ac.uk), C[3]NL, Division of Brain Sciences, Imperial College London, Hammersmith Hospital, Du Cane Road, London, W12 0NN, UK

# Abstract

In recent years, there have been many computational simulations of spontaneous neural dynamics. Here, we explore a model of spontaneous neural dynamics and allow it to control a virtual agent moving in a simple environment. This setup generates interesting brain-environment feedback interactions that rapidly destabilize neural and behavioral dynamics and suggest the need for homeostatic mechanisms. We investigate roles for both local homeostatic plasticity (local inhibition adjusting over time to balance excitatory input) as well as macroscopic "task negative" activity (that compensates for "task positive", sensory input) in regulating both neural activity and resulting behavior (trajectories through the environment). Our results suggest complementary functional roles for both local homeostatic plasticity and balanced activity across brain regions in maintaining neural and behavioral dynamics. These findings suggest important functional roles for homeostatic systems in maintaining neural and behavioral dynamics and suggest a novel functional role for frequently reported macroscopic "task-negative" patterns of activity (e.g., the default mode network).

# Introduction

In recent years, there has been increasing evidence that homeostatic systems play an important role in regulating neural activity. At the microscopic level, experimental and theoretical work suggests that the balance of local excitation and inhibition (E/I) has important computational properties [1-2]. Further, such E/I balance can be maintained with relatively simple local homeostatic inhibitory plasticity (e.g., [3-6]). At the macroscopic level, there is evidence from functional MRI that there is some level of balance in activity over regions. Networks of brain regions showing increased activity matched by other networks showing reduced activity (e.g., [7-9]). In our previous work, we have suggested that macroscopic brain networks may act to counterbalance task activation in other brain regions [10-12]. This may constitute (at the macroscopic scale) an analogue of inhibitory mechanisms and computational motifs seen at smaller scales [13].

Computational simulations are useful for understanding the functional roles of these homeostatic systems; however, computational models typically simulate the brain at rest or under constrained task settings (e.g., [14-15]). In the present work, we explore the regulatory role of homeostatic mechanisms by embodying a well-known neural simulation to control a simulated agent moving through a simple environment. This setup generates interesting brain-environment interactions that require homeostatic mechanisms to maintain rich neural and behavioral dynamics.

The computational simulation is based on the Greenberg-Hastings model [16], incorporating information about human structural connectivity [17] that has been previously shown to approximate empirical functional connectivity patterns [18]. In the model, a node was set to the ON state, with either a small random probability or if incoming activity was greater than a local threshold value (analogous to the amount of local inhibition).

In order to explore the interaction between brain and environment, we embodied the computational model in a simple environment. We began by defining an 'agent' that can move within a 2-dimensional plane, bounded by surrounding walls (see Figure 1). Within this framework, we defined a group of task-positive nodes (TP), which activated in response to simulated "sensory" input. Two pairs of bilateral nodes reacted to "visual" input from the environment to the simulated brain, and a pair reacted to "somatosensory" input: if input was detected these nodes were set to ON. A further pair of nodes simulated "motor" output from the simulated brain to the agent in the environment; their activity determined the movements of the agent.

This simple setup allows us to observe interesting interactions between the simulated neural network and the environment. Specifically, the following environment/agent closed-loop interaction occurs: (i) different parts of the environment evoke different amounts of visual and sensory stimulation, which subsequently (ii) alter regional/microscopic? neural activity, (iii) leading to altered neural dynamics macroscopically, across the entire model; (iv) these altered dynamics in turn change the motor output from the model, v) changing the agent's trajectory in the environment which in turn alters subsequent sensory input. This presents a challenge for network models, especially models focusing on spontaneous, rich dynamics, that often require careful parameterization to remain in a specific dynamic regime, and so typically are investigated in static situations (i.e., where the input to the model is stationary, such as Guassian noise). In such models, changes to the model input typically lead to destabilization of the dynamics (i.e., a shift to either random, saturated, or absent patterns of activity).

Balanced activity has been shown to facilitate a rich, spontaneous dynamical regime that is robust to different parameter values, [14,19]; this has not been explored in non-stationary scenarios, such as occur in this type of embodied models where brain-model interactions can occur. Therefore, we incorporated two mechanisms in the model to balance activity: first, a simplified version of the local (i.e., within-node) homeostatic plasticity rule presented in [3] and employed in a similar macroscopic neural model in [20]. This mechanism adjusts the local threshold (inhibition) at each node, balancing against incoming excitatory activity from other nodes, and so driving time-averaged local activity to approximate a pre-specified, small target activity rate. In addition to the local mechanism, we also employed a macroscopic balancing mechanism across brain regions, such that the activity of the task positive nodes (i.e., the six "sensory" nodes) was balanced over time by bilateral task negative nodes (TN); i.e., task negative nodes were ON but switched to off when a given sensory node was turned ON. The choice of the TN nodes was loosely based on the default mode network pattern of task-evoked relative deactivations from fMRI/PET [21], which we have previously suggested, may constitute a macroscopic balancing system [10].

Here, we explore the complementary roles that these two balancing systems may play in maintaining flexible dynamics in the embodied situation with sensory input from and motor output to the environment. In particular, we assess the agent's neural dynamics and trajectory through the environment, and demonstrate that these balancing mechanisms allow the agent to escape constrained environment-brain feedback loops, and more completely traverse the environment.

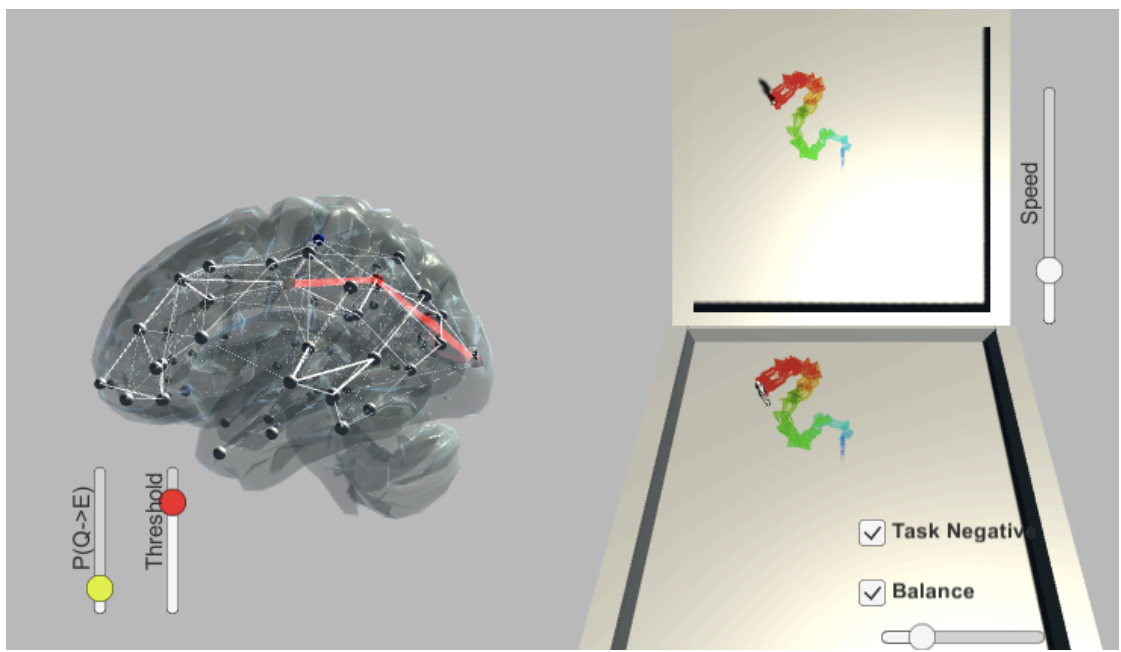

*Figure 1: Illustration of the neural model (left)* and *two perspectives of the model-controlled agent (grey figure, with its colored trail over time) placed in the environment (right). "Visual" or "somatosensory" sensory input to the agent, depends on proximity to the wall around the edge of the environment. The agent moves based activity in "motor" nodes of the model. The model*

## Results

We start by considering simulated neural and movement dynamics, without the explicit task negative mechanism. Consistent with previous results [20], we observe that over time the homeostatic model adapts the threshold weights such that time-averaged excitatory activation approximates the pre-specified target activity, $\rho = 0.1$ (Figure 2A). Moreover, the model demonstrates high levels of persistent variability around this average values (Figure 2A and 2C), even though there is relatively little intrinsic noise in the system. The model also displays non-zero but relatively weak positive correlations between nodes' activity timecourses (Figure 2D).

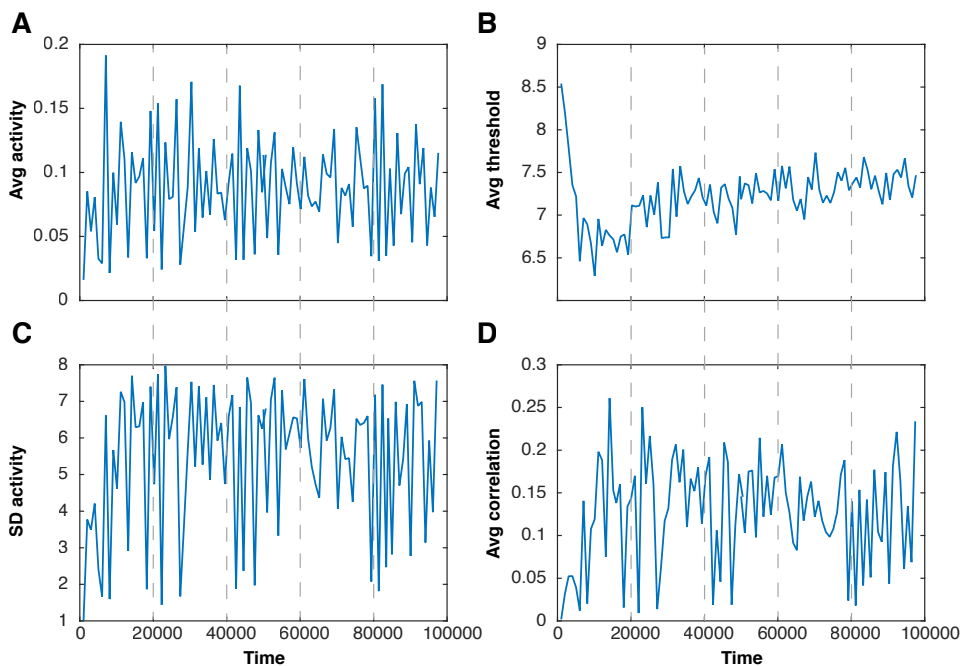

*Figure 2:* *How the simulation with local inhibitory plasticity changes over time (timecourses are of average values from 1000 epochs).* ***A****: mean activation approaches the pre-specified target value ρ=0.1.* ***B****: this is accomplished by reductions and then gradual increases in local thresholds (consistent with [20]).* ***C****: We observe increased variability of activity (standard deviation of activity over time, averaged across nodes). Finally,* ***D****: we also observe an*

*increase in connectivity across the network (measured as mean correlation between nodes). (Results are presented for a single 100,000 long training run, although qualitatively similar results were found for different initial conditions and random seeds).*

We also observed the simulated trajectory: over time the model moves into a regime with generally higher levels of movement (i.e., left/right rotation and/or forward motion) (Figure 3A), although there is considerable variability (i.e., the mean level of movement and activity varies considerably over time). Several example trajectories (each 1000 epochs long) are presented in (Figure 3B).

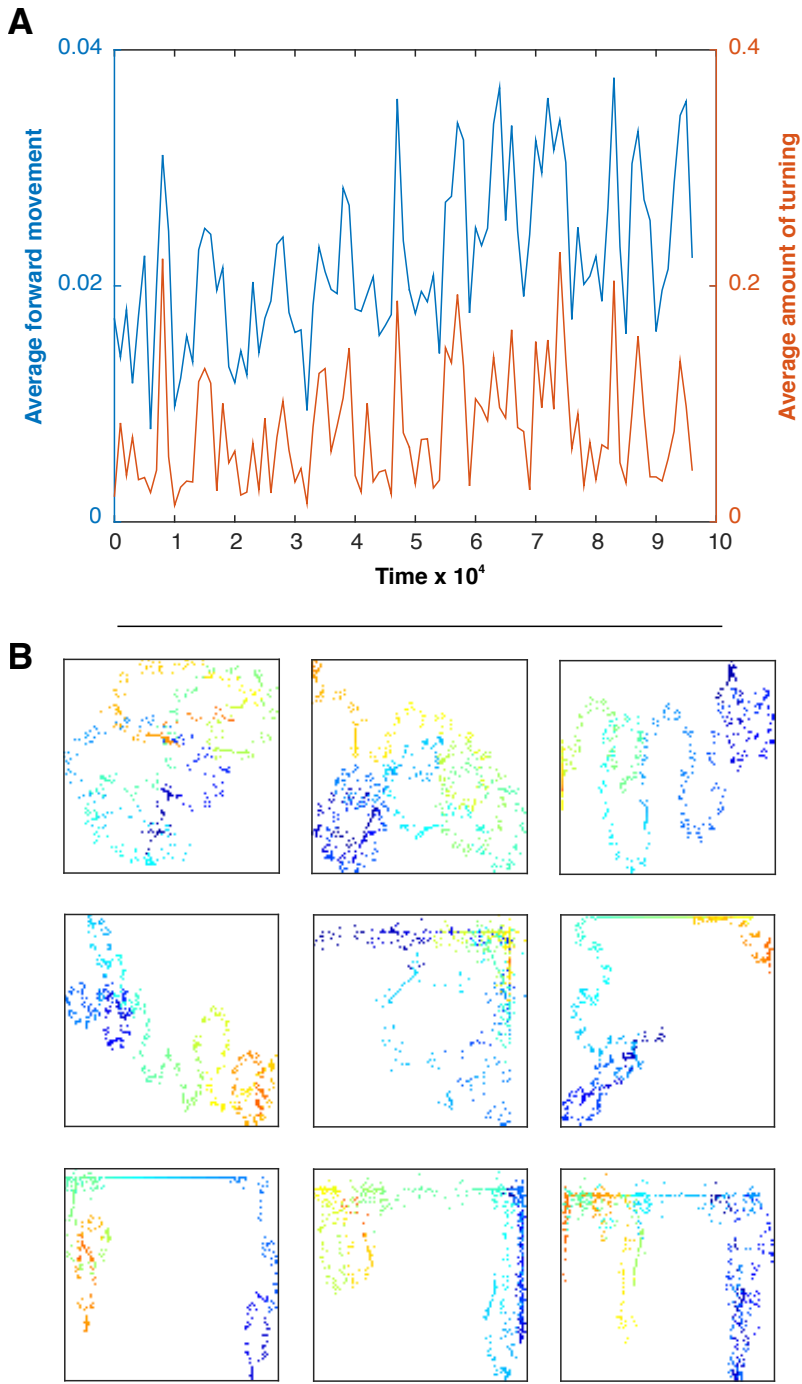

*Figure 3: A* "Motion" output (either nodes linked to "forward" output to the agent or nodes linked to "turn" over time. This is calculated from the activity of the "motor" nodes). Note that this figure presents the average output generated by the model rather than the motion of the actual agent, which can be impeded by obstacles (i.e., walls) in the environment, in this case, the simulated brain could be sending a move forward command, but this cannot be achieved because of the wall. *B* Example trajectories of the agent over

*nine randomly chosen time windows of 1000 epochs. Light colors are earlier in the time window, warm colors are later in the time window.*

However, we observe that the dynamic regimes of the simulated neural activity and motion are not stationary; the model can be characterized as alternating between periods of high and low activity, with correspondingly high and low amounts of movement. Similarly, it never settles down into a stable regime, with a single distribution of movements/activity (see Figure 4 for an illustrative shorter 1000 epoch time period). This alternating pattern reflects a feedback loop arising from how the agent interacts with the environment. The level of sensory input: i) alters the level of simulated activity, which ii) alters the level of motor output, iii) that manifests in agent movement, which in turn, iv) may alter subsequent sensory input.

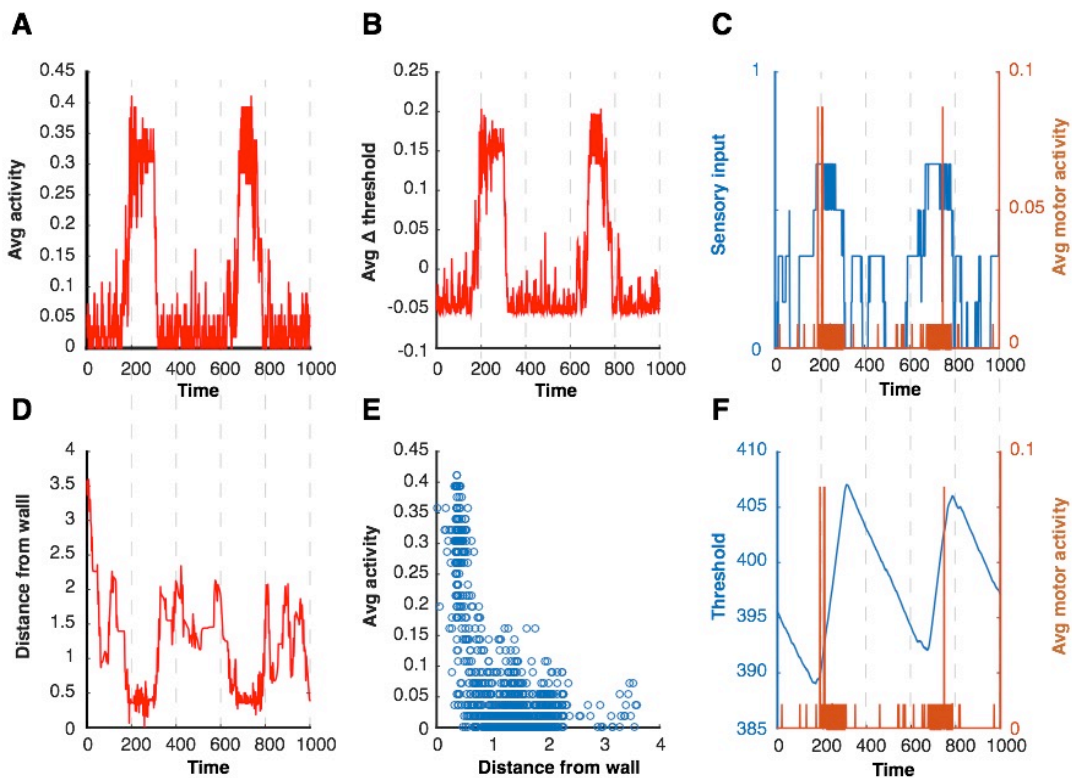

*Figure 4: An illustration of the brain/environment interaction. A: At short time periods, there are brief, but sustained periods of elevated activity substantially above the target rate. B: The homeostatic rule is continuously adapting to the level of activity, tightly matched sustained changes in the local threshold. C: Model alternating between low and high average threshold levels (or inhibition). D: The sustained changes are being driven by the sensory input*

*the model receives, which closely match overall activity; **E:** This demonstrates how sensory input and activity are related to how close the agent is to a wall (where sensory input is highest, given triggering of "visual" and "somatosensory" sensors. **F:** Strong relationship between proximity to the wall and level of activity. **G/H** the threshold change (G) and the sensory input (H) (left axes) relate to average movement (right axis), over time.*

A simple illustration of this feedback is presented in Figure 4 (and supplementary video: https://github.com/c3nl-neuraldynamics/Avatar/blob/master/Figures/FigureS1.gif; there are high levels of simulated visual and somatosensory input at the edges of the environment (that is touching and/or looking at the wall) and low or no sensory input in the center. Therefore, high sensory input triggers higher average activity which leads to elevated motor output, which on average moves the agent forward into the wall and in turn keeps the input activity high. This feedback cycle means the agent remains trying to run into the wall, trapped next to the wall and with sustained elevated activity patterns. In contrast, away from the wall, where there is little or no sensory feedback, the levels of activity remain low, and motor output and movement are either low or not present, meaning that the agent remains in a very low simulated neural activity and movement regime. The presence of homeostatic plasticity mechanism that tunes the threshold at each node to balance excitatory input from connected nodes ensures that the agent does not stay trapped in either state for long. As the threshold for activity (varying depending on the local homeostatic mechanism) at individual nodes increases (in the high activity state) or decreases (in the low activity state), the average activity level adapts to the target level. This results in the agent escaping the 'trap', with resulting activity levels closer to the target level $\rho$ and, consequently, more stable simulated neural and movement dynamics.

The model without local homeostasis is typically unable to cope with the sensory/motor feedback system. Local thresholds can be chosen to allow interesting dynamics (i.e., variable movements/neural activity); however, these must be chosen to either allow rich dynamics in the presence of sensory input (i.e., with higher local inhibition) or dynamics in the absence of sensory input (i.e., with lower local inhibition). Therefore, over time the agent will tend to either: a) remain approximately stationary in a low-sensory area with local thresholds too great to allow much exploration (i.e., near stationary, see Figure 5C); or b) become trapped in a high-sensory area running into the wall (e.g., a corner) (see Figure 5A and 5B).

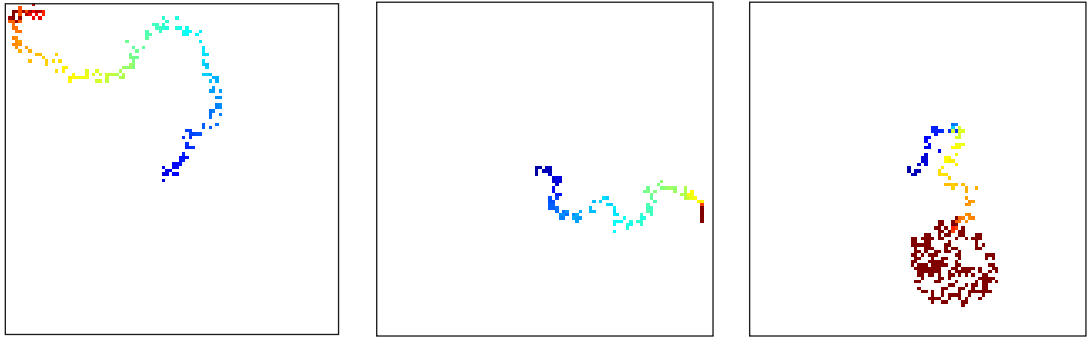

Figure 5: *Example trajectories (from initialization) of the agent without local homeostatic plasticity, and threshold weights set uniformly at 5, 7 or 9 and run for 1000 epochs. With lower weights, the agent receives high levels of excitatory activity across the brain, and walks into the wall and is trapped. With higher weights, activity within the network is very low and driven mainly by random local excitation rather than activity propagating through connections; as a result the agent moves very little over the course of the 1000 epochs. Cool colors are earlier in the time window, warm colors are later in the time window.*

While the model with local homeostasis is able to deal with this sensory-motor interaction, the addition of an explicit task-negative system, alongside the homeostatic learning rule, facilitates stable simulated neural dynamics and behavioral trajectories through the environment. This occurs because the task negative system balances changes in external input to the model so that the number of activated units (sensory nodes or task negative nodes) remains constant irrespective of interactions with the environment.

To demonstrate the complementary role of the two homeostatic systems, we compared the model with just the local homeostatic mechanism with the model with the local homeostatic mechanism and the task-negative system on a range of measures assessing the model's dynamics (simulated neural activity, simulated movement, and threshold changes: *all differences reported were significant at t>3*). We see that for both types of model, there are similar levels of mean simulated activity across the network (excluding sensory or task-negative nodes), approximately equal to the target rate (Figure 6A). However, we see that activity and threshold weight changes are less variable for the TN (Figure 6C) model. This is the case when considering both the standard deviation and the coefficient of variation (s.d./mean) of activity. Also, there is a strong negative relationship between distance from the wall and the amount of activity in the local homeostasis model (Spearman Rho=-0.45), whereas this relationship becomes much smaller in the task-negative model (Spearman Rho=-0.25), suggesting the feedback loop between brain/environment is less influential.

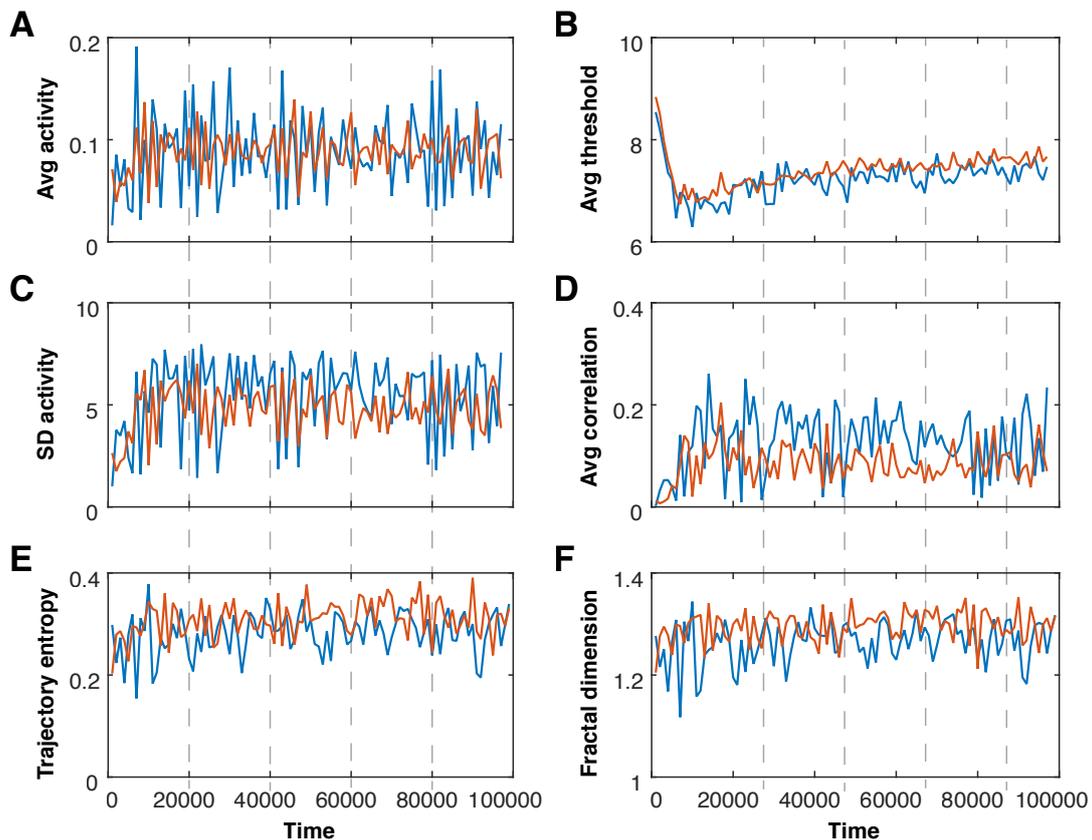

*Figure 6: Comparing simulations with and without TN (blue is without, red is with TN). We see that while the level of activation is similar in the two types of model (A, mean activity for non-TN or task positive nodes), there are higher average thresholds (B, again for non-TN or TP nodes), possibly because there are never completely quiescent periods; importantly there is*

*substantially reduced variability in activity over time for the TN model (C); and lower average correlations between nodes (D). All results presented here and in the text comparing simulations with and without TN were highly statistically significant (t>3), with data sampled from each of the 1000 epoch blocks. Also, (E) image entropy from plotted trajectories and (F) fractal dimension from plotted trajectories calculated over 1000 epoch blocks.*

When considering motor output (i.e., "motor signals" sent to control movement of the agent, e.g., turn left, go forward), we observe that there is significantly higher entropy for the plotted trajectories of the TN model. Similarly, when observing the movement of the agent we see that there is more movement in general (Figure 7A,B), and that the path of the agent has a higher fractal dimension and higher entropy (taking the 2d entropy of the image of the path over 1000 epochs) for the TN model (and has visited significantly more of all possible locations, see Figure 7C), suggesting it enjoys a more complex pattern covering more of the environment and is less affected by the feedback system.

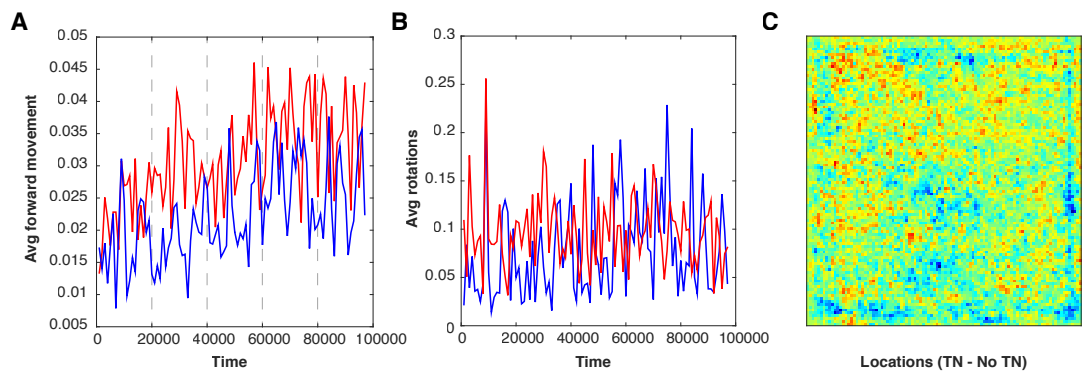

*Figure 7: A and B (left and center) present forward motion (i.e., average distance moved per epoch), and absolute average turning over time for both the TN (red) and non TN (blue) simulations. There is generally more movement for the TN model. C, (right), location of the agent averaged across the last 50,000 epochs contrasting the TN model with the non-TN model (smoothed with a Gaussian kernel, sigma=0.5). We see that in general the TN model visits more of the environment than the non-TN (i.e., pixels with warm colors, indicate TN > non-TN; cold colors are the reverse).*

The location of the sensory input systems. motor output systems task-negative nodes were chosen relatively arbitrarily. This is because the coarse resolution of the parcellation means that assigning sensory or motor labels to nodes is inherently very approximate. As such, we do not wish to draw parallels with specific brain regions or networks (e.g., from the functional imaging literature). However, it is interesting to look at the relationships between task positive and task negative nodes and how this affects the neural simulations. Therefore, we repeated the simulations, randomly varying the location of the task negative nodes. We ran 65 simulations that were the identical to the simulations detailed previously except: 1) they were shorter (5000 epochs, for practical reasons); 2) the TN were randomly chosen from any of the nodes that had not been defined as "sensory" or "motor". We observed that the stability of the model is dependent on the choice of the individual TN nodes. Specifically, we see that the solution is more stable when TP are linked by more walks to TN nodes (calculated using the Brain Connectivity Toolbox [22]). This was assessed by counting the number of walks (of <6 steps) between TP and TN nodes. This resulted in distance measures between somatosensory TP-TN nodes, and between visual TP-TN nodes. These distances were entered into general linear models to predict measures of neural dynamics. This analysis revealed that the standard deviation and coefficients of variation for threshold weights and activation as well as the correlation between distance to wall and activation were all significantly related to distance between TP-TN ($p<0.01$; $F(3,63)>4.5$). Therefore, as TP-TN nodes became more closely connected, the model finds a more stable solution, and neural dynamics are less affected by brain-environment feedback.

## Discussion

This model is unequivocally not intended to be a detailed model of all aspects of real embodied cognition or of actual neural and sensorimotor systems; instead, in both regards, it is highly simplified. We acknowledge that there have been many arbitrary design choices, and do not intend this to be a definite presentation of how to model brain/environment/behavior interactions. Such interactions systems are likely to be far more complex, possibly non-stationary, and will depend on the complexity both of the neural system, but also the complexity of the motor and sensory systems. Instead, the example we present here is a useful toy example; the simplification allows us to consider the interactions between macroscopic brain networks, neural dynamics and the environment to better understand possible functional roles of homeostatic systems in the brain.

With the above as a strong caveat, however, our findings highlight the challenges that feedback between environment and brain presents to neural models. Further, our results suggest that modeling spontaneous dynamics at rest (e.g., [23-25]) or with a simple task such as encoding a sensory stimulus (e.g., [14-15]) is different to modeling sensori-motor interactions with an environment; further, the existence of a closed-loop feedback made the roles of homeostatic mechanisms more important and obvious. In our case, we observed that without the local homeostatic plasticity, the agent in the environment would become trapped in either a stationary state (with high levels of local inhibition) or would be in a permanent state of motion (with too little local inhibition). Instead, we observe that plasticity is a constant feature of the system. Initially, there are large changes in local thresholds across time points, as the model approximately balances average incoming excitation at each node. As time progresses, however, the weight changes become smaller, but never drop completely to zero.

We also observed that local homeostatic plasticity could be complemented by adding a macroscopic, task-negative system to compensate for sensory-induced activity, across the whole simulated brain. The simple system we implemented, modeled on patterns of task negative deactivation from the fMRI/PET literature (e.g., [10]) counteracted the destabilizing effects from the changing amount of sensory input that the model receives in different locations in the environment. Without the task negative system, the overall level of activity within the model is more dependent on the level of sensory input (i.e., "touching" or "seeing" the wall). This makes the task of the local homeostatic plasticity mechanism harder, since exogenous input to the system varies considerably. Instead, the task negative system simply balances the level of exogenous activity to a constant amount, such that task negative input decreases as sensory input increases and visa versa. This means that the environment/brain feedback loop does not change the overall level of incoming activity to the model, therefore facilitating the local homeostatic plasticity to find a more stable solution, i.e., one that requires the smallest weight changes to approximate the target activation rate. Further, what we observe are different balancing systems operating at different spatial and temporal scales and with different specific mechanisms. This is consistent with the proposed description of normalization found in many neural systems [26], which provides a canonical computation across scales and implementations, and results in improved neural coding efficiency and sensitivity.

From a traditional cognitive neuroscience perspective, this way of thinking about task negative systems may sit somewhat uncomfortably. What we have been describing as task negative may provide a partial functional explanation for the default mode network. The default mode network is a well-characterized, frequently observed and relatively poorly understood macroscopic brain network located in areas of the brain not associated with sensorimotor activity; the default mode network has been observed across ontogeny [27], phylogeny [28], and found across different cognitive and sensorimotor tasks [29] and implicated through abnormal function in many disorders [30]. According to our findings, the default mode network can be thought of as acting as a counterweight, or as an endogenous generator of neural activity that allows the neural system to remain relatively stable in an inherently unstable world. One analogy could be with the vascular system of warm-blooded animals, which attempts to maintain a constant body temperature, irrespective of the temperature outside, in order to maintain a stable environment for chemical reactions to take place, ultimately allowing more flexible behavior. (We note that the proposed balancing functional role for task negative brain networks does not preclude more traditional cognitive roles ascribed to them, such as internal mentation. We hypothesise that task negative systems could have initially evolved to perform some basic neural function, such as balancing incoming sensory activity, and eventually been exaptively repurposed over evolution to perform more specific cognitive functions that occur when external input is not present).

Following this explanation of the task negative systems in general and the default mode network more specifically, we see that task-negative systems may not strictly be "necessary" for accomplishing any task. As such, lesioning task negative regions is unlikely to disturb any associated function entirely, and as such task negative systems may appear to be epiphenomenal. However, just as a sailing boat does not require a keel to move (the keel counterbalances the forces on the sail, facilitating stability and allowing a wider range of movement and greater speed), the brain may have a greater range of neural state and potentially be more controllable, when it is properly counterbalanced. It might only be over longer time periods when initially adapting to a novel environment or across development that damage to task negative systems becomes particularly disabling, failing to facilitate other adaptive systems as efficiently.

We find that the model finds a more stable solution if TN nodes are strongly linked to task positive nodes. This is consistent with the presence of multiple TN systems in the brain ([9,31]) rather than a single TN. This would be consistent with the brain being configured to involve active counterbalancing systems, such that in the optimal case each task positive configuration would have a matched task negative one, to balance it. However, we acknowledge that there are likely to be trade-offs between having a perfectly balanced system and having a functional one. In our model, the nodes do not carry out any actual computations and are assigned unitary roles (in terms of sensorimotor function), which is unlikely to be true; both of these, and other (e.g., onto- or phylogenetic) considerations could affect the type of TN that evolution has arrived at.

Finally, in order to achieve a relatively stable solution with rich spontaneous dynamics and interactions with the environment, the system may have to encode (in the local inhibitory weights) information about the world, and the agent's movement in it. Given the relative simplicity of the environment in the current simulation, the presence of local thresholds is adequate to facilitate a relatively stable solution. However, as the environment (and sensory input systems) becomes more complex, it will be necessary to use more sophisticated models with more flexibility. If the repertoire of brain states is to be more fully explored in the face of this increasing complexity, then it will be necessary to capture more information about the environment/sensory systems. This leaves open questions about the roles of other types of learning (e.g., longer-distance excitatory and reinforcement learning) and their roles in supporting the system staying in a rich dynamical regime in a complex environment with complex sensorimotor systems and with more cognitive control mechanism.

## Methods

### Empirical Structural Connectivity

Simulated activity patterns were generated from a computational model constrained by empirical measures of white-matter structural connectivity between 66 cortical regions of the human brain, defined by diffusion tensor imaging (DTI) [17]. This structural network has been used in a range of previous computational models to demonstrate emergent properties of resting state functional connectivity [14,20,24,32]. A full methodology, describing the generation of this matrix $\langle C \rangle$ is available in [17]. In brief: measures of length and strength of stream-line based connectivity were estimated using Deterministic tractography of DSI datasets (TR=4.2s, TE=89s, 129 gradient directions max b-value 9000s/mm$^2$) of the brain in 5 healthy control subjects. A high-dimensional ROI based connectivity approach was projected though the 66 regions of the Desikan-Killianey atlas (FreeSurfer http://surfer.nmr.mgh.harvard.edu/), such that $C_{i,j}$ is the number of streamlines connecting nodes $i$ and $j$.

### Computational Model

*Neural Dynamics*

To simulate brain activity, we defined a simple model based on the Greenberg-Hastings model, which has been shown in previous work to approximate patterns of empirical functional connectivity [18]. At each time point, *t*, each node, *i*, in the model can be in one of three states, $S_{i,t}$ : excitatory (E), quiescent (Q), or refractory (R). Nodes changed state according the following simple probabilities: $p_i(E \rightarrow R) = 1$; $p_i(R \rightarrow Q) = 1$; $p_i(Q \rightarrow E) = 10^{-1}$. Importantly, nodes would also change from Q->E if the summed input from *n* connected nodes, *j*, was greater than a local threshold value: $\sum_{j=1}^{n} C_{i,j} S_{j,t-1} > T_i$. The strength of the activation threshold, $T_i$, could be tuned to separately at each node (see below). $S_{i,t}$ was binarized so that E was coded as 1, R or Q as 0.

## Homeostatic plasticity

For most of the simulations, we used a local homeostatic plasticity mechanism as follows: we allowed the activation threshold to vary by a small amount based on the activity in each node at the previous time-step, according to the following rule similar (but simplified) to that introduced in [3] and used in [20] and with a similar (but simpler) effect of balancing incoming excitation from connected nodes:

$$\delta t_i = \alpha(S_{i,t} - \rho)$$

where $\rho$ is a target activation and $\alpha$ is a learning rate. Thus in the case that the activity of $i$ is 1, and $\rho < 1$ there is an increase in the threshold whereas, otherwise the threshold decreases. Therefore, the time-averaged activity of $S_i$, will approximate $\rho$.

## Environmental Embedding

The motor activity (movement) of the agent was defined by two commands; Turn ($h$) in radians per update step and Move ($v$) which moved the agent forward v world units. The activity within these two parameters at each time-step was determined by the simulated neural activity at four nodes (two rotate and two advance nodes) of the computational model. We chose these nodes to be bilaterally symmetrical such that they approximately correspond to motor related regions in the brain (n.b., we make no claims that this anatomical correspondence is correct or that the results are dependent on this). If a rotate node was active, the agent would attempt to turn ≈30° in that direction. If both nodes were active, then the effect would cancel out this out. If a single forward node was active, the agent would move forward 1/10 of a unit the arbitrary world space, if both forward nodes were active, the unit would move forward 1 unit of world space. In addition, we added some temporal smoothing across time for activity within the move such that the move command described was 7/8 of the activity of the relevant assigned node, and 1/8 of the activity of the previous time step. (The amount of this smoothing and the values of how nodes translated into movement were chosen semi-arbitrarily, to produce agent motion that appeared superficially plausible, i.e., neither very fast or slow).

Sensory information ('visual' perception) from the environment was integrated into the computational model through the use of two horizontal ray-traces emanating from each "eye" of the agent and offset by ±10° from the vertical. A distance threshold was defined, such that if an object (i.e., the wall, in this simple environment) was less <2 units of world space then a specific node ("near visual") of the model was set to the E state, if an object was detected between 2 and 10 world units away then the "far visual" node was set to excitatory. In addition, to "visual" input, we also defined a rudimentary "somatosensory" input, whereby if model had collided with any other object in the environment then specific "somatosensory" nodes for collisions on either of the Left or Right side of the agent within the computational model were set to the E state.

### Task-Negative nodes

In order to explore the effect of balance between task positive (TP) and task negative (TN) networks, we defined for some simulations, a collection of TN nodes that were anti-correlated with the TP nodes described above. These TN nodes were defined as two (bilateral) pairs of task negative nodes approximately corresponding to regions that consistently show relative deactivation across many empirical fMRI tasks were chosen (although, this was still a relatively arbitrary decision and we do not wish to make any claims based on anatomical precision). These nodes were set to the E state if TN nodes (i.e., the "visual", or "somatosensory" nodes were in the Q or R states, and Q, when the TN nodes were activated, such that TP and TN nodes were anti-correlated.

Further, given that TN activity is task specific [10,33,34] we defined two TN nodes (one on the left and its homologous region on the right) that were set to be excitatory when "visual" activity was not excitatory; and, a separate pair of nodes (again bilateral, homologous) were set to be activated when "somatosensory" activity was silent. For most simulations, the location of the task negative nodes were kept constant. However, in an additional set of simulations, the task negative nodes were randomly re-positioned by picking random bilateral homologous pairs of nodes from the network. Unless stated otherwise, the results presented are from a single model run for 100,000 epochs. However, we repeated the model two further times with different random seeds (so different patterns of excitatory noise, resulting simulated dynamics and movements), replicating the results presented below.

## Acknowledgements and Supplementary Materials

The code/implementation (in Unity) and compiled versions of the model are available to at https://github.com/c3nl-neuraldynamics/Avatar/releases. Special thanks go to Eva Papaeliopoulos for getting us started in making computer games.